\documentclass[journal]{IEEEtran}
\usepackage{amsmath,amssymb}
\usepackage{psfrag}
\usepackage{epsfig}
\usepackage{cite}
\usepackage{graphics}
\usepackage{color}
\usepackage{subfigure}
\usepackage{multirow}
\usepackage{threeparttable}
\usepackage{booktabs}
\usepackage{url}
\usepackage{mathrsfs}
\begin{document}
\title{From Simplicity to Complexity Based on Consensus: A Case Study}
\author{Yinyan Zhang and
Shuai Li, \IEEEmembership{Member,~IEEE}
\thanks{ This work is supported by the National Natural Science Foundation of
China (with number 61401385),
by Hong Kong Research Grants Council Early Career Scheme (with
number 25214015), and also by Departmental General Research Fund of
Hong Kong Polytechnic University (with number G.61.37.UA7L).}
\thanks{Y. Zhang and S. Li are with the Department of Computing,
The Hong Kong Polytechnic University, Hung Hom, Kowloon, Hong Kong,
China (e-mails: yinyan.zhang@connect.polyu.hk; shuaili@polyu.edu.hk).
}
}
\markboth{IEEE TRANSACTIONS ON AUTOMATIC CONTROL}%
{Shell \MakeLowercase{\textit{et al.}}: Bare Demo of IEEEtran.cls
for Journals}
\maketitle
\begin{abstract}
Distributed consensus has been intensively studied in recent years
as a means to mitigate state differences among dynamic nodes on a
graph. It has been successfully employed in various applications,
e.g., formation control of multi-robots, load balancing,  clock
synchronization. However, almost all existing applications cast an
impression of consensus as a simple process to iteratively reach
agreement, without any clue on possibility to generate advanced
complexity, say shortest path planning, which has been proved to be
NP-hard. Counter-intuitively, we show for the first time that the
complexity of shortest path planning can emerge from a perturbed
version of min-consensus protocol, which as a case study may shed
lights to researchers in the field of distributed control to
re-think the nature of complexity and the distance between control
and intelligence. Besides, we rigorously prove the convergence of
graph dynamics and its equivalence to shortest path solutions. An
illustrative simulation on a small scale graph is provided to show
the convergence of the biased min-consensus dynamics to shortest
path solution over the graph. To demonstrate the scalability to
large scale problems, a graph with 43826 nodes, which corresponds to
a map of a maze in 2D, is considered in the simulation study. Apart
from possible applications in robot path planning, the result is
further extended to robot complete coverage, showing its potential
in real practice such as cleaning robots.
\end{abstract}



\section{Introduction}
\IEEEPARstart{C}{complex} network attracts a lot of attention from
the control community. The consensus problem is a fundamental
problem in this area. Consensus means that a network of nodes
reaches an agreement on certain quantities of interest through
information exchanging between neighbors. Consensus provides a
fundamental rule to reach global agreement through local
interactions and has been successfully employed to design
distributed algorithms for various applications, e.g., clock
synchronization \cite{ncsbo} and multi-robot formation control
\cite{dcafm,dacfc}.

The recent decades have witnessed the development of consensus on
distributed graphs. Olfati-Saber and Murray \cite{cpino} proposed a
general framework for solving the consensus problem of graph nodes
with single-integrator dynamics under fixed or switching topologies
and communication delays. In terms of the consensus of nodes with
double-integrator dynamics, necessary and sufficient conditions were
proposed in \cite{snasc,nascf}. The average consensus of dynamic
graphs described by high-order integrators was addressed in
\cite{acoho}. In \cite{comsa}, a unified viewpoint was proposed for
the consensus of multi-agent systems and the synchronization of
complex networks. In recent years, results have also been reported
to deal with challenging issues on consensus under different
scenarios, such as communication delay \cite{conms,cciun,citnw},
noises \cite{dcais}, uncertainty of node dynamics \cite{coaco} and
switching topology \cite{scino}. Recently, under a unified
framework, Cao {\it et al.} \cite{ftcpc} proposed a finite-time
convergent distributed consensus algorithm to address the consensus
of node dynamics with partially known nonlinearity in a
state-dependent interaction graph. Many results are reported about
the applications of consensus to specific systems
\cite{tvfcf,cpfom,mtcba}. For example, Dong {\it et al.}
\cite{tvfcf} adopted consensus protocols to deal with a time-varying
formation control problem of unmanned aerial vehicles. Yang {\it et
al.} \cite{mtcba} presented minimum-time consensus-based distributed
algorithms to handle load shedding and economic dispatch of power
systems.

There is a clear gap between the consensus of a dynamic graph and
the problem of finding the shortest path over a graph. Consensus is
a simple evolution while path planning (especially shortest path
planning) is a complex behavior. Different from the consensus
problem, path planning problems are fundamental in artificial
intelligence, which are related to complex decision-making
processes, and some of them have been proved to be NP hard
\cite{trspp,ahsfp,rdoan}. Among them, the classical shortest path
planning problem is about finding a shortest path from a given
source position to a destination position, which is a complex
combinational optimization problem. Note that there may exist
multiple shortest paths. A shortest path problem becomes more
complex if there are multiple possible destination positions, which
means that we do not know what the end of the shortest path should
be before we solve the problem. Current, commonly used algorithms
for shortest path planning include A$^*$ algorithm, Dijkstra' s
algorithm, and their numerous variants \cite{abbgs}. To our
knowledge, there is no existing results that bring the two disjoint
problems, i.e., consensus and shortest path planning, together and
seek solutions for the latter one by means of consensus.

Under mild conditions, graph nodes running consensus protocols
recursively converge to an agreement in their state values
\cite{cpino}. Following intuition, it is reasonable to expect an
approximate agreement in the state values of graph nodes when
perturbing a consensus protocol with a bias. Surprisingly, our
finding reveals that the result is far beyond a simple approximate
agreement and it has direct correspondence to shortest path of the
graph. This finding gives us a positive answer to the fundament
question: whether distributed consensus, as a simple evolution can
generate advanced complexity, e.g., shortest path planning. To the
best of the authors' knowledge, there is no existing result on
solving shortest path planning from the perspective of consensus.
Further investigation of this problem may trigger our attempt to
re-think the nature of intelligence and complexity, and develop
tractable ways to address advanced complexity by using tools from
the field of distributed control.

The rest of this paper is organized as follows. In Section
\ref{sec.2}, we present a brief review on consensus and graph
theory. In Section \ref{sec.3}, a biased min-consensus protocol is
established by perturbing the existing min-consensus protocol and
the corresponding theoretical analysis on the stability of the
biased min-consensus protocol is also presented. In Section
\ref{sec.4}, we reveal the equivalence between the result of biased
min-consensus and shortest path planning. In Section \ref{sec.5},
simulations and corresponding discussions are presented to further
substantiate the efficacy of the biased min-consensus protocol in
solving shortest path planning problems and its potential in various
applications, e.g., maze solving and complete coverage. In Section
\ref{sec.6}, we conclude this paper with final remarks. Before
ending this introductory section, the main contributions of this
paper are listed as follows.
\begin{itemize}
\item[1)] This paper shows that consensus as a simple evolution can generate a complex
behavior implying advanced intelligence (i.e., shortest path
planning which is NP hard), indicating the potential on
investigating problems arising in artificial intelligence from the
perspective of distributed control.
\item[2)] The emergence of complexity, i.e., the shortest path solution on a graph, from biased min-consensus is theoretically analyzed and proved rigorously.
\item[3)] Apart from finding the shortest path on a graph, we show applications on using biased min-consensus for maze solving and complete coverage, which demonstrates application potentials of biased min-consensus.
\end{itemize}

\section{Background}\label{sec.2}

In the section, we briefly overview the background and review basics
about graph theory and consensus.

\subsection{Graph Theory}

The graph theory is a useful tool for investigations on consensus of
network of nodes. We only present the definitions necessary for this
paper. Definitions on directed graphs, Laplacian matrices and
spanning trees, which are also widely adopted in consensus-related
researches, can be found in literature \cite{Bookagt}.

Let $\mathbb{G} = (\mathbb{V}, \mathbb{E})$ denote an undirected
connected graph with the set of vertices (or nodes) denoted by
$\mathbb{V}=\{1,2,\cdots,n\}$ and the set of edges denoted by
$\mathbb{E}$. The value of node $i$ in the graph is denoted by
$x_i$. The edge connecting node $i$ and node $j$ is denoted by
$(i;j)$ with $i = 1, 2,\cdots,n$ and $j = 1, 2,\cdots,n$, where $n$
denotes the number of vertices in the graph. The set of neighbors of
node $i$ is denoted by $\mathbb{N}(i)= \{j\mid(i;j)\in\mathbb{E}\}$.
The weight of edge $(i;j)$ in a undirected graph is denoted by
$w_{ij}$. Specifically, if edge $(i;j)$ exists, then $w_{ij}>0$;
otherwise, $w_{ij}=0$.

For the convenience of latter illustration, we present the
definitions of the shortest path problem in the graph theory as
follows.

{\it Definition 1} (Shortest path problem): The shortest path
problem defined in graph $\mathbb{G}=(\mathbb{V},\mathbb{E})$ is to
find a path from a node $s\in\mathbb{V}$ to another node
$v\in\mathbb{V}$ such that the sum of the weights of its constituent
edges is minimized.


\subsection{Consensus}

In this paper, we only consider the situation that the communication
is bidirectional, which corresponds to an undirected connected
graph. A general definition of consensus is the $\chi$ consensus,
which is presented as follows.

{\it Definition 2} ($\chi$ consensus \cite{cpfno}): Consider a
network consisting of $n$ nodes defined in undirected connected
graph $\mathbb{G}=(\mathbb{V},\mathbb{E})$ with the state value of
node $i$ denoted by $x_i$. We say that nodes asymptotically achieves
$\chi$ consensus if
$\lim_{t\rightarrow+\infty}x_i=\chi(\mathbf{x}(0))$, $\forall
i\in\mathbb{V}$, where
$\mathbf{x}(0)=[x_1(0),x_2(0),\cdots,x_n(0)]^\text{T}\in\mathbb{R}^n$
denotes the initial state of all the nodes and
$\chi(\mathbf{x}):\mathbb{R}^n\rightarrow\mathbb{R}$ denotes a
function for which the function value is unique for any argument
$\mathbf{x}$.

Other types of consensus may be viewed as special cases of the
$\chi$ consensus, which mainly includes the min-consensus, the
max-consensus and the average consensus. In the following
subsections, we briefly show two protocols about the the
min-consensus and the average consensus.

\subsection{Min-Consensus}

Consider a network consisting of $n$ nodes defined in undirected
connected graph $\mathbb{G}=(\mathbb{V},\mathbb{E})$ consisting of
$n$ nodes. The min-consensus is such that the multi-agent achieves
$\chi$ consensus with
$\chi(\mathbf{x})=\min\{x_1(0),x_2(0),\cdots,x_n(0)\}$ \cite{cpfno}.

In a general network, there are two types of nodes, i.e., leader
nodes and follower nodes. Let $\mathbb{S}_1$ and $\mathbb{S}_2$
denote sets of leader nodes and follower nodes respectively. For a
network with static leader nodes, an intuitive distributed
min-consensus protocol is \cite{dpfdw}:
\begin{equation}\label{equ.min}
\begin{cases}
\dot{x}_i=0,~i\in\mathbb{S}_1,\\
\dot{x}_i=-x_i+\min_{j\in\mathbb{N}(i)}\{x_j\}, ~i\in\mathbb{S}_2.
\end{cases}
\end{equation}
The min-consensus is similar to the max-consensus. Some recent
results about the min-consensus or the max-consensus can be found in
\cite{amcpw,aomca,socbt}.

\subsection{Average Consensus}

Different from the min-consensus, the average consensus is such that
the nodes in a network achieves $\chi$ consensus with
$\chi(\mathbf{x})=\sum^n_{i=1}x_i(0)/n$ \cite{cpfno}. A classical
leaderless average consensus protocol is
\begin{equation*}
\dot{x}_i=\sum_{j\in\mathbb{N}(i)}w_{ij}(x_i-x_j).
\end{equation*}
Most of the existing results on consensus are about the average
consensus (see \cite{ccoma,ebbfm,dsfac} for example).

\section{Biased Min-Consensus}\label{sec.3}
In this section, we perturb the existing min-consensus protocol to
establish biased min-consensus. Then, we analyze its properties and
stability.

\subsection{Protocol}

In min-consensus protocol (\ref{equ.min}), the leader nodes do not
receive information from other nodes, they are static with the
time-derivatives being 0. Meanwhile, the follower nodes are dynamic.
They receive information $x_j$ from their neighbors via
communication. In terms of the issues arising from communication, as
mentioned in the Introduction part, some of the existing results
concerns communication delay. Now we consider another case, where
the information neighbor node $j$ that follower $i$ receives is
$x_j+w_{ij}$, which yields the following biased min-consensus
protocol:
\begin{equation}\label{nn.biased}
\begin{cases}
\dot{x}_i=0,~i\in\mathbb{S}_1,\\
\varepsilon\dot{x}_i=-x_i+\text{min}_{j\in\mathbb{N}(i)}\{x_j+w_{ij}\},~
i\in\mathbb{S}_2
\end{cases}
\end{equation}
where constant $\varepsilon>0\in\mathbb{R}$ is a protocol parameter.

We have the following remarks on an intuitive explanation and the
distributed property of the proposed biased min-consensus protocol.

{\it Remark 1:} an intuitive explanation for biased min-consensus
protocol (\ref{nn.biased}) is as follows. The term
$-x_i+\text{min}_{j\in\mathbb{N}(i)}\{x_j+w_{ij}\}$ means that the
changes of state values of follower nodes are a feedback result of
the differences between their state values and information from
their neighbors. In addition, parameter $\varepsilon$ can be viewed
as a gain to adjust the strength of feedback.

{\it Remark 2:} The biased min-consensus protocol is distributed
since each node either only receive information from its neighbors
or does not receive any information from its neighbors. The former
corresponds to follower nodes and the latter corresponds to leader
nodes. For example, follower node $i$ receives information
$x_j+w_{ij}$ from each neighbor defined in set $\mathbb{N}(i)$.
Then, only the minimum value of $x_j+w_{ij}$ have an impact on the
state value of node $i$.

\subsection{Properties}\label{sec.3.2}

The analysis on non-biased consensus of multi-node network defined
on undirected graphs, the definition of Laplacian matrix and its
properties are often adopted. However, due to the existence of
biased term $w_{ij}$ in biased min-consensus protocol
(\ref{nn.biased}), traditional analysis for consensus does not
apply. In this section, we present theoretical analysis on the
stability of biased min-consensus protocol (\ref{nn.biased}).

For the convenience of illustration, we denote the right-hand side
of biased min-consensus protocol (\ref{nn.biased}) by $e_i$.
Specifically, $e_i=0$ for $i\in\mathbb{S}_1$ and
$e_i=\min_{j\in\mathbb{N}(i)}\{x_j+w_{ij}\}-x_i$ for
$i\in\mathbb{S}_2$. We define the upper bound of $e_i$ as
\begin{equation}\label{equ.uppere}
\bar{e}=\max_{i\in\mathbb{V}}\{{e}_i\},
\end{equation} with the corresponding node set denoted by
\begin{equation}\label{equ.upperset}
\bar{\mathbb{S}}=\text{arg}~\max_{i\in\mathbb{V}}\{e_i\}.
\end{equation}
Meanwhile, we define the lower bound of $e_i$ as
\begin{equation}\label{equ.lowere}
\underline{e}=\min_{i\in\mathbb{V}}\{{e}_i\}
\end{equation}
and the corresponding node set is denoted by
\begin{equation}\label{equ.lowerset}
\underline{\mathbb{S}}=\text{arg}~\min_{i\in\mathbb{V}}\{e_i\}.
\end{equation}
Let $\emptyset$ denote the empty set which does not contain any
element. The parent node set of node $i$ is defined as follows:
\begin{equation}\label{equ.parent}
\begin{cases}
\mathbb{P}(i)=\emptyset,~i\in\mathbb{S}_1,\\
\mathbb{P}(i)=\text{arg}~\min_{j\in\mathbb{N}(i)}\{x_j+w_{ij}\},~i\in\mathbb{S}_2.
\end{cases}
\end{equation}
Since we only
consider connected undirected graphs in this paper,
$\mathbb{P}(i)\neq\emptyset$, $\forall i\in\mathbb{S}_2$. Similarly,
we can also define the child node set of node $i$ as $\mathbb{C}(i)$
with $\mathbb{C}_i=\{k\in\mathbb{V}\mid i\in\mathbb{P}(k)\}$.

Now we are ready to present properties of biased min-consensus
protocol (\ref{nn.biased}). We first consider upper bound
$\bar{e}_i$ and lower bound $\underline{e}_i$ of the right-hand side
of biased min-consensus protocol (\ref{nn.biased}). The two
quantities are global information, which show how the overall
multi-node network evolutes with time.

{\it Lemma 1:} Upper bound $\bar{e}_i$ defined in (\ref{equ.uppere})
for biased min-consensus protocol (\ref{nn.biased}) is monotonically
non-increasing.

{\it Proof:} In light of biased min-consensus protocol
(\ref{nn.biased}), for $i\in\mathbb{S}_2$, we have
$\dot{e}_i=\sum_{j\in\mathbb{P}(i)}\lambda_j\dot{x}_j-\dot{x}_i$
with $0<\lambda\leq1$ and $\sum_{j\in\mathbb{P}(i)}\lambda_j=1$. It
follows that, for $i\in\mathbb{S}_2$,
\begin{equation*}
\begin{aligned}
\dot{e}_i&=\sum_{j\in\mathbb{P}(i)}\lambda_j\dot{x}_j-\sum_{j\in\mathbb{P}(i)}\lambda_j\dot{x}_i\\
&=\sum_{j\in\mathbb{P}(i)}\lambda_j(\dot{x}_j-\dot{x}_i)\\
&=\sum_{j\in\mathbb{P}(i)}\frac{\lambda_j}{\varepsilon}(e_j-e_i).
\end{aligned}
\end{equation*}
Besides, for $i\in\mathbb{S}_1$, we have $\dot{e}_i=0$. Accordingly,
for $\bar{e}$, we have
$\dot{\bar{e}}=\sum_{i\in\bar{\mathbb{S}}}\delta_i\dot{e}_i$ with
$0\leq\delta_i\leq1$ and $\sum_{i\in\bar{\mathbb{S}}}\delta_i=1$.
Divide set $\bar{\mathbb{S}}$ into two subsets:
$\bar{\mathbb{S}}=\bar{\mathbb{S}}\cap\mathbb{V}=\bar{\mathbb{S}}\cap(\mathbb{S}_1+\mathbb{S}_2)
=(\bar{\mathbb{S}}\cap\mathbb{S}_1)+(\bar{\mathbb{S}}\cap\mathbb{S}_2)$.
It follows that
$\dot{\bar{e}}=\sum_{i\in\bar{\mathbb{S}}\cap\mathbb{S}_1}\delta_i\dot{e}_i+
\sum_{i\in\bar{\mathbb{S}}\cap\mathbb{S}_2}\delta_i\dot{e}_i$. Note
that
$\sum_{i\in\bar{\mathbb{S}}\cap\mathbb{S}_1}\delta_i\dot{e}_i=0$
since $\dot{e}_i=0$ for $i\in\mathbb{S}_1$. Then, we have
\begin{equation}\label{fath}
\begin{aligned}
\dot{\bar{e}}=\sum_{i\in\bar{\mathbb{S}}\cap\mathbb{S}_2}\delta_i\dot{e}_i
=\sum_{i\in\bar{\mathbb{S}}\cap\mathbb{S}_2}\sum_{j\in\mathbb{P}(i)}\frac{\lambda_i}{\varepsilon}\delta_j
(e_j-e_i).
\end{aligned}
\end{equation}
Since $i\in\bar{\mathbb{S}}$, we have $e_i=\bar{e}\geq e_j$, i.e.,
$e_j-e_i\leq0$, $\forall j\in\mathbb{V}$. Recall that
$\lambda_i\geq0$, $\delta_j\geq0$ and $\varepsilon>0$. Then, we have
$\dot{\bar{e}}\leq0$. In other words, $\bar{e}$ is monotonically
non-increasing. The proof is complete. $\hfill\Box$

{\it Lemma 2:} Lower bound $\underline{e}_i$ defined in
(\ref{equ.lowere}) for biased min-consensus protocol
(\ref{nn.biased}) is monotonically non-decreasing.

{\it Proof:} It can be generalized from the proof of Lemma 1 and is
thus omitted. $\hfill\Box$

Based on Lemma 1 and Lemma 2, we also have the following two lemmas
about biased consensus protocol (\ref{nn.biased}).

{\it Lemma 3:} When $t\rightarrow+\infty$,
$\mathbb{P}(i)\subset\bar{\mathbb{S}}$, $\forall
i\in\bar{\mathbb{S}}$ with $\mathbb{P}(i)$ and $\bar{\mathbb{S}}$
defined in (\ref{equ.parent}) and (\ref{equ.upperset}),
respectively, for biased consensus protocol (\ref{nn.biased}).

{\it Proof:} According to Lemma 1, $\dot{\bar{e}}\leq0$, i.e., upper
$\bar{e}$ (\ref{equ.uppere}) is monotonically non-increasing.
According to Lemma 2, $\dot{\underline{e}}\geq0$, i.e., lower bound
$\underline{e}_i$ (\ref{equ.lowere}) is monotonically
non-decreasing. It follows that
$\lim_{t\rightarrow+\infty}{\dot{\bar{e}}}=0$. From equation
(\ref{fath}), we further have $e_j(t)=e_i(t)$,
$t\rightarrow+\infty$, $j\in\mathbb{P}(i)$, $\forall
i\in\bar{\mathbb{S}}$. In other words,
$\mathbb{P}(i)\subset\bar{\mathbb{S}}$, $\forall
i\in\bar{\mathbb{S}}$ when $t\rightarrow+\infty$. The proof is
complete. $\hfill\Box$

{\it Lemma 4:} For sets $\mathbb{P}(i)$ and $\underline{\mathbb{S}}$
defined in (\ref{equ.parent}) and (\ref{equ.lowerset}),
respectively, when $t\rightarrow+\infty$,
$\mathbb{P}(i)\subset\underline{\mathbb{S}}$, $\forall
i\in\underline{\mathbb{S}}$.

{\it Proof:} The proof is similar to that of Lemma 3 and thus
omitted. $\hfill\Box$

Note that Lemma 1 and Lemma 2 imply that $\bar{e}_i$ and
$\underline{e}_i$ converge to constants with time. However, it
remains unknown whether both of them converge to 0. Therefore, it is
necessary to analyze whether state value $x_i$ becomes unbounded
with time. To address this issue, we have the follow lemma.

{\it Lemma 5:} Consider an undirected connected graph $\mathbb{G}$.
State value $x_i$ is upper bounded, $\forall t>0$, $\forall
i\in\mathbb{V}$, for biased min-consensus protocol
(\ref{nn.biased}).

{\it Proof:} According to Lemma 1, $\dot{\bar{e}}\leq0$, i.e.,
$\bar{e}(t)=\max_{i\in\mathbb{V}}\{-x_i(t)+\min_{j\in\mathbb{N}(i)}\{x_j(t)+w_{ij}\}\}\leq\bar{e}(0)$,
$\forall t>0$, $\forall i\in\mathbb{V}$. It follows that
$x_k+w_{ik}-x_i\leq\bar{e}(0)$, $\forall k\in\mathbb{P}(i)$,
$\forall i\in\bar{\mathbb{S}}$. According to Theorem 2,
$\dot{\underline{e}}\geq0$, $\forall t>0$. Together with
$\dot{\bar{e}}\leq0$, we further have $\underline{e}(0)\leq
e_i(t)\leq\bar{e}(0)$, $\forall t>0$. Recall that
$e_i=-x_i+\min_{j\in\mathbb{N}(i)}\{x_j+w_{ij}\}=-x_i+x_k+w_{ik}$
with $k\in\mathbb{P}(i)$. Then, $\underline{e}(0)\leq
-x_i+x_k+w_{ik}\leq\bar{e}(0)$, i.e.,
$x_i\leq-\underline{e}(0)+x_k+w_{ik}$ with $k\in\mathbb{P}(i)$. From
the definition of $\mathbb{P}(i)$, we have $x_k+w_{ik}\leq
x_j+w_{ij}$, $\forall j\in\mathbb{N}(i)$, $\forall
k\in\mathbb{P}(i)$. It follows that
\begin{equation}\label{equ.rec}
x_i\leq-\underline{e}(0)+x_j+w_{ij},~\forall j\in\mathbb{N}(i).
\end{equation}
As we assume that the graph is undirected and connected, we can
always find a path from a node $i_1\in\mathbb{S}_1$ to a node
$i_\eta\in\mathbb{S}_2$. Suppose that the path consists of $\eta$
($\eta\geq2$) nodes including node $i_1$ and node $i_\eta$.
Considering that $\dot{x}_i=0$ for $i\in\mathbb{S}_1$, from
inequality (\ref{equ.rec}), we have $x_{i_n}\leq
-\underline{e}(0)(\eta-1)+\max_{i\in\mathbb{S}_1}\{x_i(0)\}+(\eta-1)\max_{(i;j)\in\mathbb{E}}\{w_{ij}\}$.
It follows that $x_i\leq
-\underline{e}(0)(\eta-1)+\max_{i\in\mathbb{S}_1}\{x_i(0)\}+(\eta-1)\max_{(i;j)\in\mathbb{E}}\{w_{ij}\}$,
$i\in\mathbb{V}$. The proof is complete. $\hfill\Box$

Now we consider the relationship between set $\mathbb{S}_1$ and set
$\underline{\mathbb{S}}$ when $t\rightarrow+\infty$, which is
presented in the following lemma.

{\it Lemma 6:} Consider an undirected connected graph $\mathbb{G}$.
When $t\rightarrow+\infty$,
$\underline{\mathbb{S}}\cap\mathbb{S}_1\neq\emptyset$ for biased
min-consensus protocol (\ref{nn.biased}) with set
$\underline{\mathbb{S}}$ defined in (\ref{equ.lowere}) and
$\mathbb{S}_1$ denoting the set of leader nodes.

{\it Proof:} From Lemma 1 and Lemma 2 and the definitions of
$\bar{e}$ and $\underline{e}$, we have $\bar{e}(0)\geq\bar{e}(t)\geq
e(t)\geq\underline{e}(t)\geq\underline{e}(0)$ and
$\dot{\underline{e}}(t)\geq0$ for all $t\geq0$. It follows that
$\lim_{t\rightarrow+\infty}\dot{\underline{e}}(t)=0$, which
indicates $e_i(t)$ equals for all $i\in\underline{\mathbb{S}}$ and
$e_i\leq0$, i.e.,
$-x_i+\min_{j\in\mathbb{N}(i)}\{x_j+w_{ij}\}\leq0$, when
$t\rightarrow+\infty$. Suppose
$\underline{\mathbb{S}}\cap\mathbb{S}_1=\emptyset$ when
$t\rightarrow+\infty$. It follows that $\mathbb{P}(i)\neq\emptyset$,
$\forall i\in\underline{\mathbb{S}}$, according to the definition of
$\mathbb{P}(i)$ in equation (\ref{equ.parent}) and the assumption
that the graph is undirected and connected. Besides, according to
Lemma 4, $\mathbb{P}(i)\subset\underline{\mathbb{S}}$, $\forall
i\in\underline{\mathbb{S}}$. Then, we have
\begin{equation}\label{equ.the4}
x_i(t)\geq\min_{j\in\mathbb{N}(i)}\{x_j(t)+w_{ij}\}>x_k(t),~\forall
k\in\mathbb{P}(i)\subset\underline{\mathbb{S}},~\forall
i\in\underline{\mathbb{S}}~
\end{equation}
when $t\rightarrow+\infty$.  Let
$x_\text{m}=\lim_{t\rightarrow+\infty}\min_{i\in\underline{\mathbb{S}}}\{x_i(t)\}$.
Then, we have
\begin{equation}\label{equ.thetem}
x_\text{m}\leq x_k(t), \forall k\in\underline{\mathbb{S}},
\end{equation}
when $t\rightarrow+\infty$. From inequality (\ref{equ.the4}), when
$t\rightarrow+\infty$, we have $x_i(t)>x_k(t)$,
$k\in\mathbb{N}(i)\subset\underline{\mathbb{S}}$, which contradicts
with inequality (\ref{equ.the4}). Therefore,
$\underline{\mathbb{S}}\cap\mathbb{S}_1\neq\emptyset$ when
$t\rightarrow+\infty$. The proof is complete. $\hfill\Box$

\subsection{Stability}

Based on the properties derived in Section \ref{sec.3.2} on bias-min
consensus (\ref{nn.biased}), we are ready to present the theorem
about the stability of (\ref{nn.biased}).

{\it Theorem 1:} Let $\mathbb{G}$ be an undirected connected graph
and suppose each node of $\mathbb{G}$ applies biased min-consensus
(\ref{nn.biased}). Then, all nodes of the graph globally and
asymptotically converge to the equilibrium point of
(\ref{nn.biased}).

{\it Proof:} Equilibrium point $\mathbf{x}^*$ of biased
min-consensus protocol (\ref{nn.biased}) satisfies the following
equation:
\begin{equation}\label{equ.stable}
\begin{cases}
x^*_i=x_i(0),~i\in\mathbb{S}_1,\\
x^*_i=\min_{j\in\mathbb{N}(i)}\{x_j+w_{ij}\},~i\in\mathbb{S}_2.
\end{cases}
\end{equation}

From Lemma 1 and Lemma 2, we have
$\lim_{t\rightarrow+\infty}\bar{e}(t)=c_0$ and
$\lim_{t\rightarrow+\infty}\underline{e}(t)=c_1$ with $c_0$ and
$c_1$ being two constants. Given that the graph is undirected and
connected, from Lemma 6,
$\underline{\mathbb{S}}\cap\mathbb{S}_1\neq\emptyset$ when
$t\rightarrow+\infty$. It follows that there exists an
$i\in(\mathbb{S}_1\cap\underline{\mathbb{S}})$. Since $e_j=0$ for
all $j\in\mathbb{S}_1$, in light of the definition of
$\underline{e}$ in equation (\ref{equ.lowere}), we further have
$\lim_{t\rightarrow+\infty}\underline{e}(t)=0$.

In light of the definition of $\bar{e}$ in equation
(\ref{equ.uppere}), we have $\bar{e}\geq0$. For
$i\in\bar{\mathbb{S}}$, from bias-min consensus protocol
(\ref{nn.biased}), we have $\varepsilon\dot{x}_i=e_i=\bar{e}\geq0$.
Note that we have proved
$\lim_{t\rightarrow+\infty}\underline{e}(t)=0$. Then, we have
$e_i\geq0$ when $t\rightarrow+\infty$. If follows that
$\varepsilon\sum_{i\in\mathbb{V}}\dot{x}_i\geq|\bar{\mathbb{S}}|\lim_{t\rightarrow+\infty}\bar{e}(t)$,
where $|\bar{\mathbb{S}}|$ denotes the number of nodes in set
$\bar{\mathbb{S}}$. Evidently, $\sum_{i\in\mathbb{V}}\dot{x}_i$ will
grow unboundedly if $\lim_{t\rightarrow+\infty}\bar{e}(t)>0$, which
contradicts with Lemma 5 (i.e., $x_i$ is bounded). Therefore,
$\lim_{t\rightarrow+\infty}\bar{e}(t)=0$.

Summarizing the above proof, one has
$\lim_{t\rightarrow+\infty}{e}_i(t)=0$. It follows that
$\lim_{t\rightarrow+\infty}\dot{x}_i(t)=0$, $\forall
i\in\mathbb{V}$. Therefore, equilibrium point $\mathbf{x}^*$ of
biased min-consensus protocol is globally stable. The proof is
complete. $\hfill\Box$

\section{Equivalence to  Shortest Path Planning}\label{sec.4}

Consensus is a simple evolution while shortest path planning is a
complex behavior which is related to high-level intelligence.
Traditionally, there is a clear gap between counter intuitions. In
this section, we present the relationship between consensus and
shortest path planning via biased min-consensus protocol
(\ref{nn.biased}).

The relationship between consensus of multi-node network and
shortest path planning defined in undirected connected graphs can be
constructed as follows. The state value of a node is the length of a
path from this node to one of the destination nodes. The destination
nodes are the static nodes with state values always being 0, and the
follower nodes correspond to the source nodes. In other words, the
set of source nodes corresponds to $\mathbb{S}_1$ and the set of
destination nodes corresponds to $\mathbb{S}_2$. Besides, if there
is a edge between two nodes, the corresponding nodes can communicate
with each other. The length of the edge connecting node $i$ and node
$j$ is denoted by $w_{ij}$. Then, we employee biased min-consensus
protocol (\ref{nn.biased}) for the nodes to communicate with their
neighbor nodes.

The following theorem shows that the equilibrium of biased
min-consensus protocol (\ref{nn.biased}) actually forms shortest
paths from any source node to destination nodes.

{\it Theorem 2:}  If $x_i(0)=0$ for all $i\in\mathbb{S}_1$, then the
equilibrium of biased min-consensus protocol (\ref{nn.biased}) forms
a solution to the corresponding shortest path problem.

{\it Proof:} According to Theorem 1, the following equilibrium of
$\mathbf{x}^*$ of biased min-consensus protocol (\ref{nn.biased}) is
asymptotically stable and satisfies equation (\ref{equ.stable}).
According to the optimality principle of Bellman's dynamic
programming \cite{oarp}, the solution of the considered shortest
path problem satisfies the following nonlinear equations:
\begin{equation}\label{equ.bell}
\begin{cases}
x_i=0,~i\in\mathbb{S}_1,\\
x_i=\min_{j\in\mathbb{N}(i)}\{x_j+w_{ij}\},~i\in\mathbb{S}_2,
\end{cases}
\end{equation}
and the solution of the nonlinear equations is unique if there
exists a shortest path. Evidently, the equilibrium of biased
min-consensus protocol (\ref{nn.biased}) satisfies nonlinear
equations (\ref{equ.bell}) if $x_i(0)=0$ for all $i\in\mathbb{S}_1$.
Therefore, the equilibrium of biased min-consensus protocol
(\ref{nn.biased}) forms a solution to the shortest path problem,
given that $x_i(0)=0$ for all $i\in\mathbb{S}_1$. The proof is
complete. $\hfill\Box$

{\it Remark 3:} Once the state values of the nodes converge to the
solution of nonlinear equation (\ref{equ.bell}), the shortest path
can be found by recursively finding the parent nodes. A series of
parent nodes forms a shortest path.

\section{Simulations And Applications}\label{sec.5}
In this section, simulations and applications (including maze
solving and complete coverage) are shown and discussed to
substantiate the efficacy of biased min-consensus (\ref{nn.biased})
for shortest path planning and validate theoretical results,
indicating potential investigations on artificial problems from the
perspective of control.

\begin{figure}[t]\centering
\includegraphics[scale=0.3]{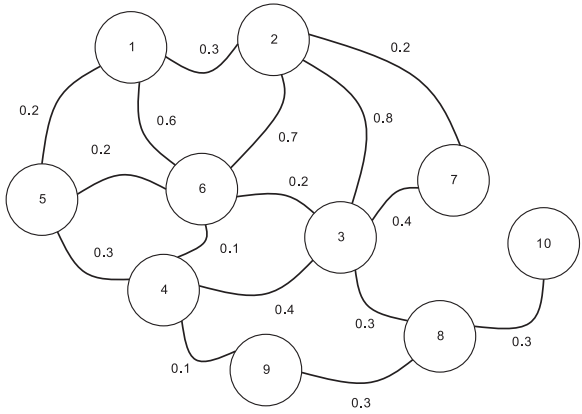}
\caption{A 10-node graph with node 1 being the destination node.
\label{fig.ex11}}
\end{figure}
\begin{figure}[t]\centering
\psfrag{i}[c][c][0.6]{$i$} \psfrag{t}[c][c][0.6]{$t$ (s)}
\psfrag{xi}[c][c][0.6]{$x_i$}
\subfigure[]{\includegraphics[scale=0.295]{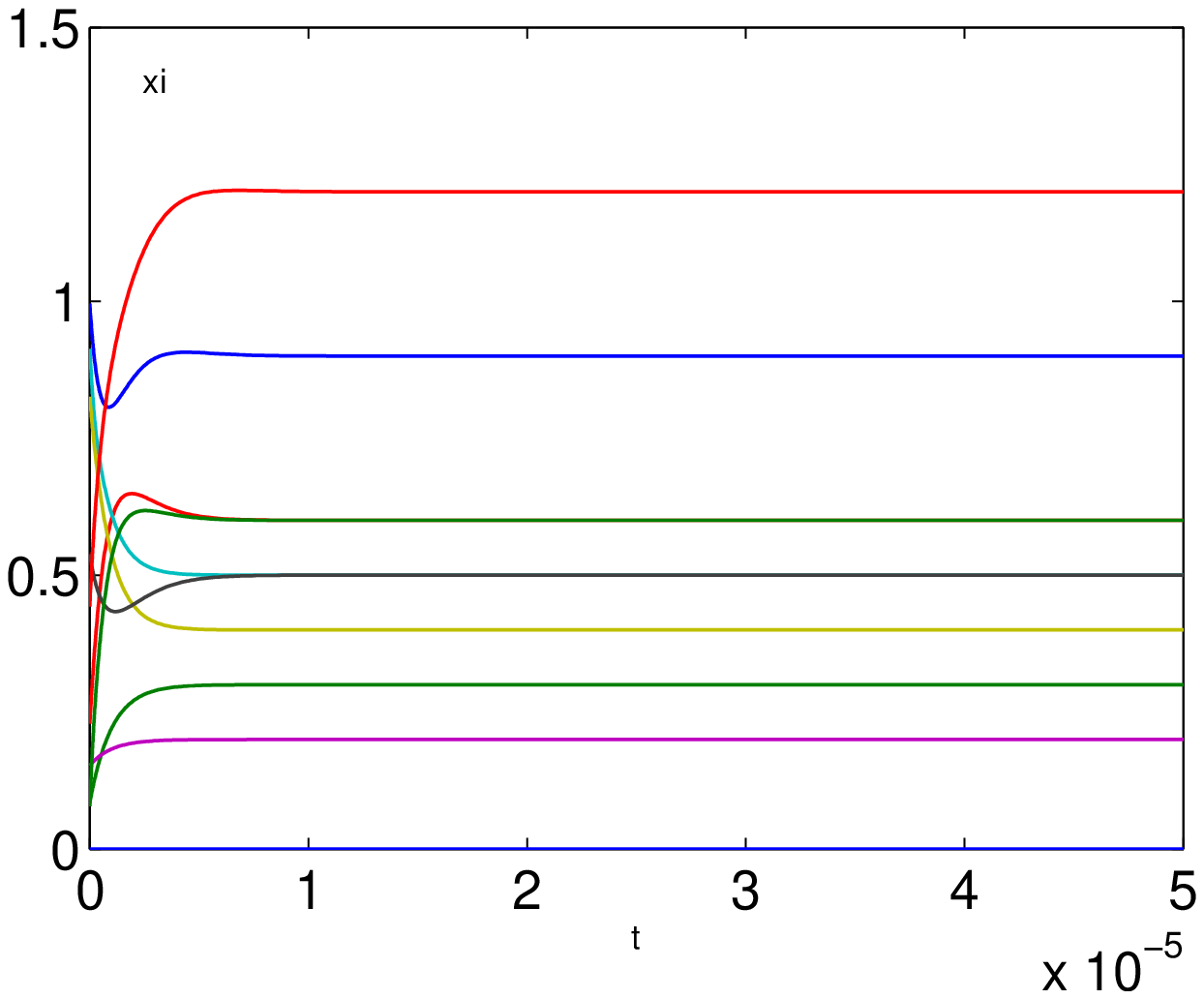}}
\subfigure[]{\includegraphics[scale=0.295]{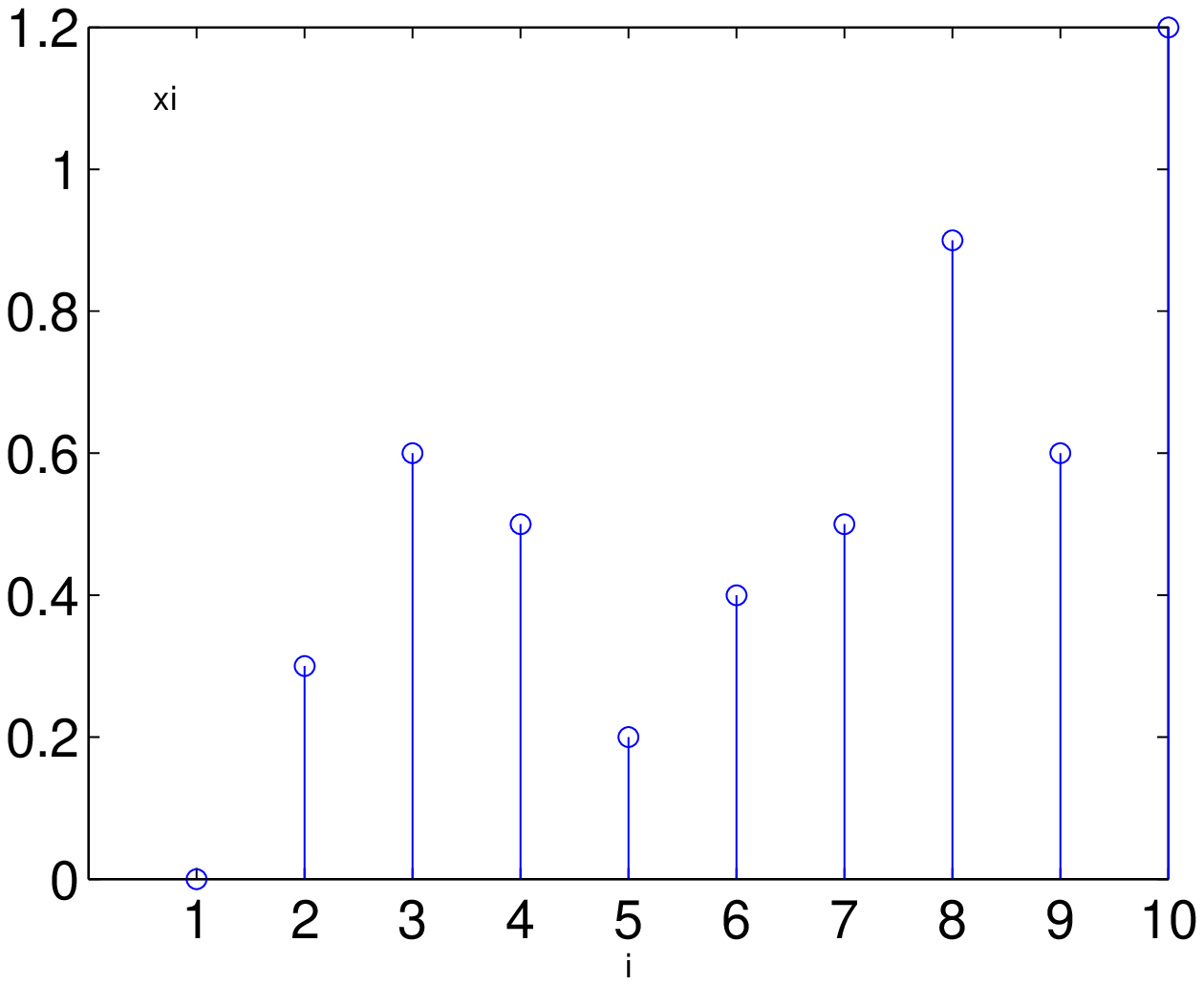}}
\caption{Time profile of biased min-consensus protocol
(\ref{nn.biased}). (a) Transient behavior of $x_i$. (b)Steady state
of $x_i$. As verified by calculation, the steady state value of
biased min-consensus is identical to the optimal distance from the
destination node as computed by Dijkstra's algorithm, and can form
the shortest path by following Remark 3. \label{fig.ex12}}
\end{figure}
\begin{figure}[t]\centering
\includegraphics[scale=0.2]{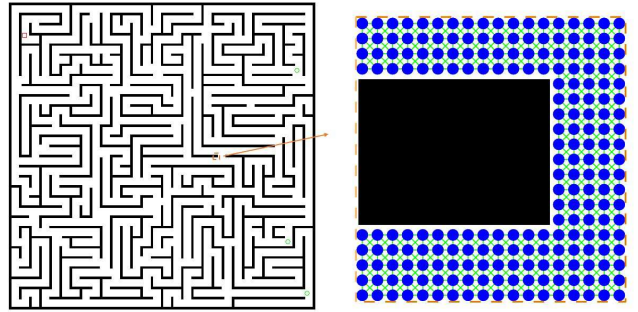}
\caption{Using biased min-consensus for maze solving. Maze graph
containing $254\times254$ pixels with the initial position marked by
a rectangle and destination positions marked by circles. A graph is
constructed by associating each free pixel that is not occupied by
obstacles with a node, and forming graph edges between nodes mapped
from neighboring pixels. This forms a large-scale connected graph
totally with 43826 nodes. \label{fig.ex21}}
\end{figure}

\subsection{Illustrative Example with A 10-Node Graph}
In this subsection, we consider the shortest path planning defined
in a 10-node graph shown in Fig. \ref{fig.ex11}. In the graph, node
1 is the destination node. With $\varepsilon=10^{-6}$, the
simulation results based on biased min-consensus protocol
(\ref{nn.biased}) are shown in Fig. \ref{fig.ex12}. As seen from
Fig. \ref{fig.ex12}(a), the min-consensus protocol is convergent.
From \ref{fig.ex12}(b), it can be readily verified that the steady
state values are the lengths of the shortest paths from the nodes to
destination node 1. According to Remark 3, we can readily find the
shortest path from any source nodes to node 1. For example, there
are two shortest path from node 10 to the destination node (i.e.,
node 1), i.e.,
$10\rightarrow8\rightarrow3\rightarrow6\rightarrow5\rightarrow1$ and
$10\rightarrow8\rightarrow9\rightarrow4\rightarrow6\rightarrow5\rightarrow1$
The results substantiate the efficacy of the biased min-consensus
protocol for solving shortest path problem and validate the
theoretical results.

\begin{figure*}[t]\centering
\subfigure[$t=0$ s]{\includegraphics[scale=0.3]{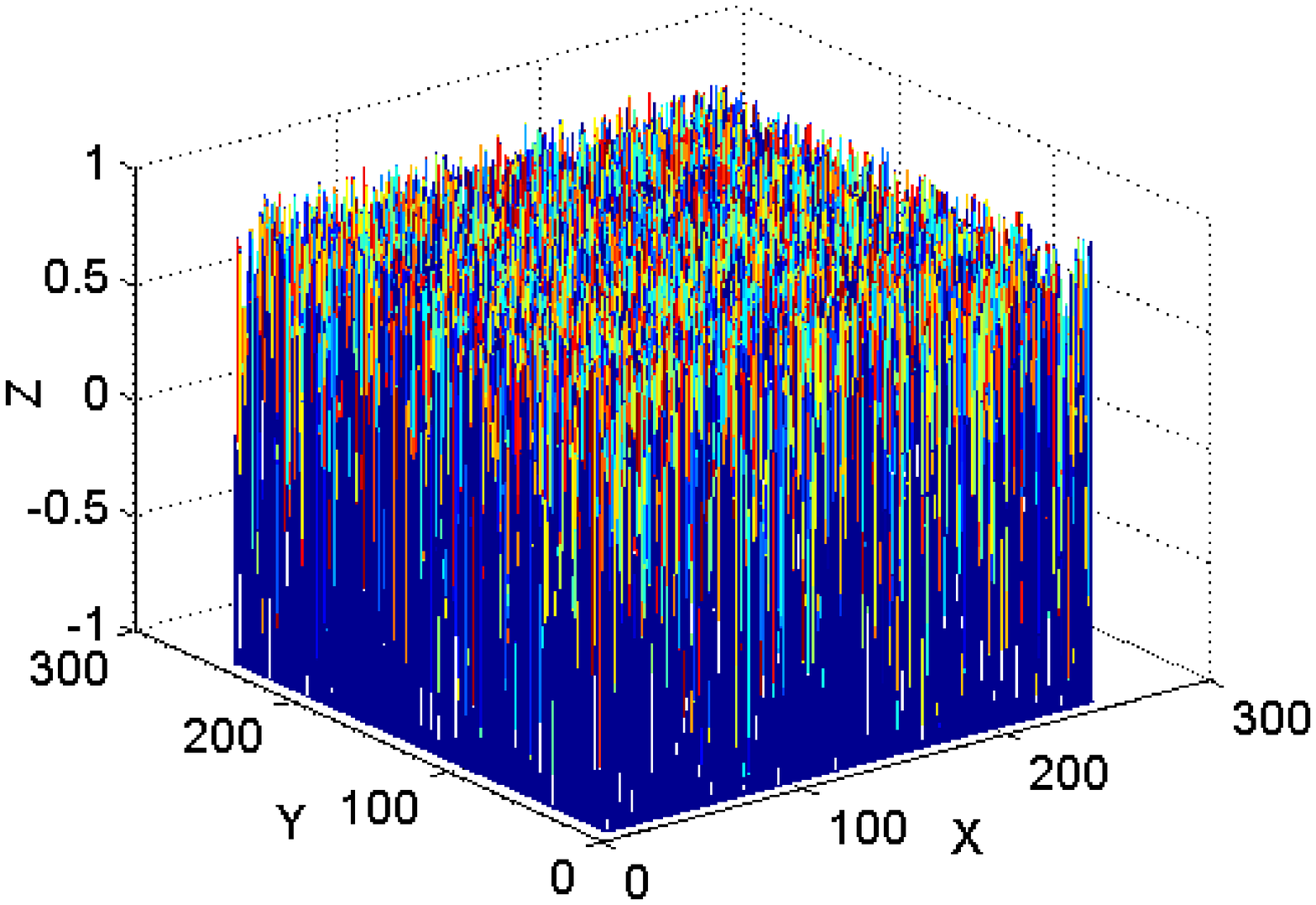}}
\subfigure[$t=0.01$ s]{\includegraphics[scale=0.3]{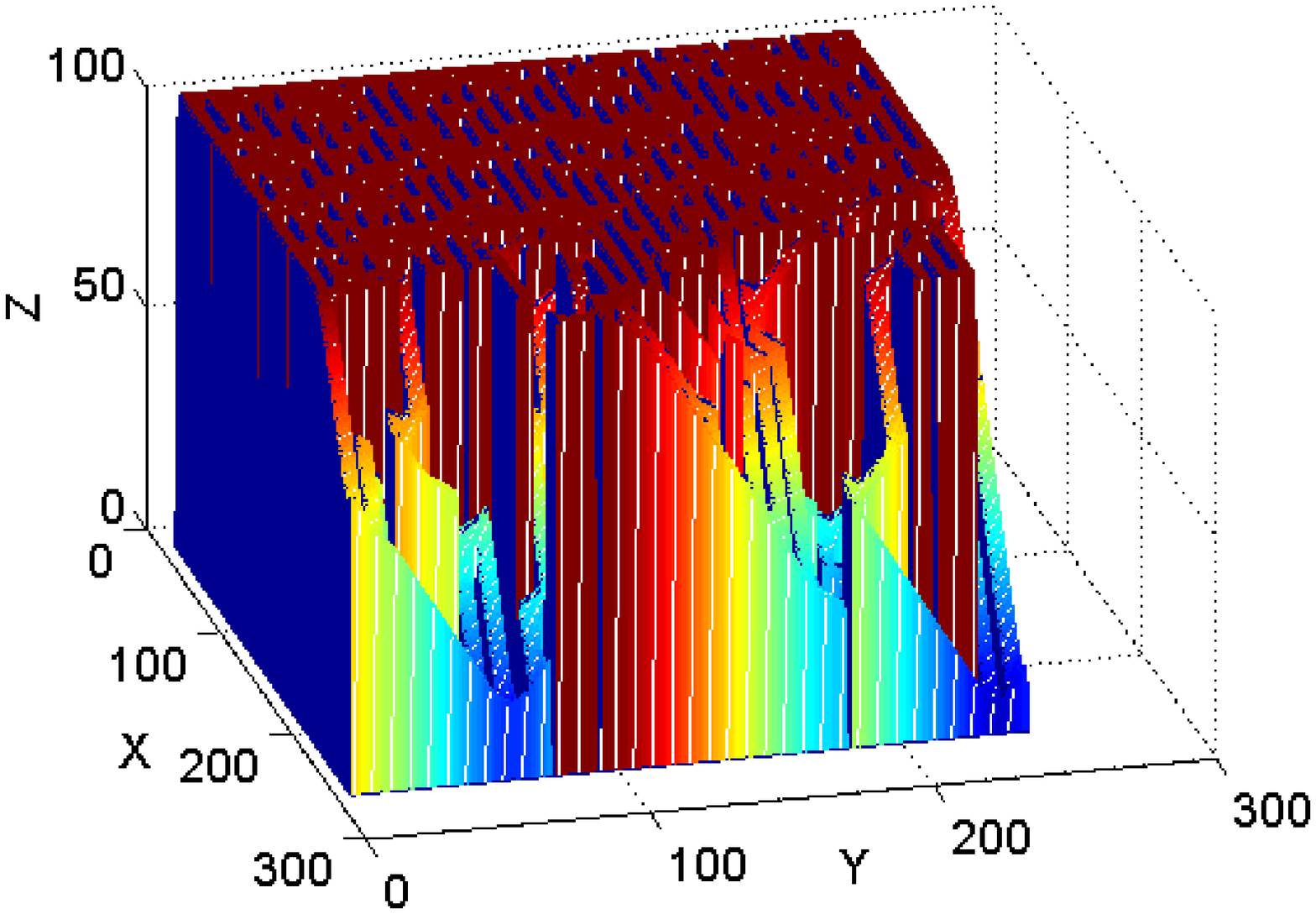}}
\subfigure[$t=0.02$ s]{\includegraphics[scale=0.3]{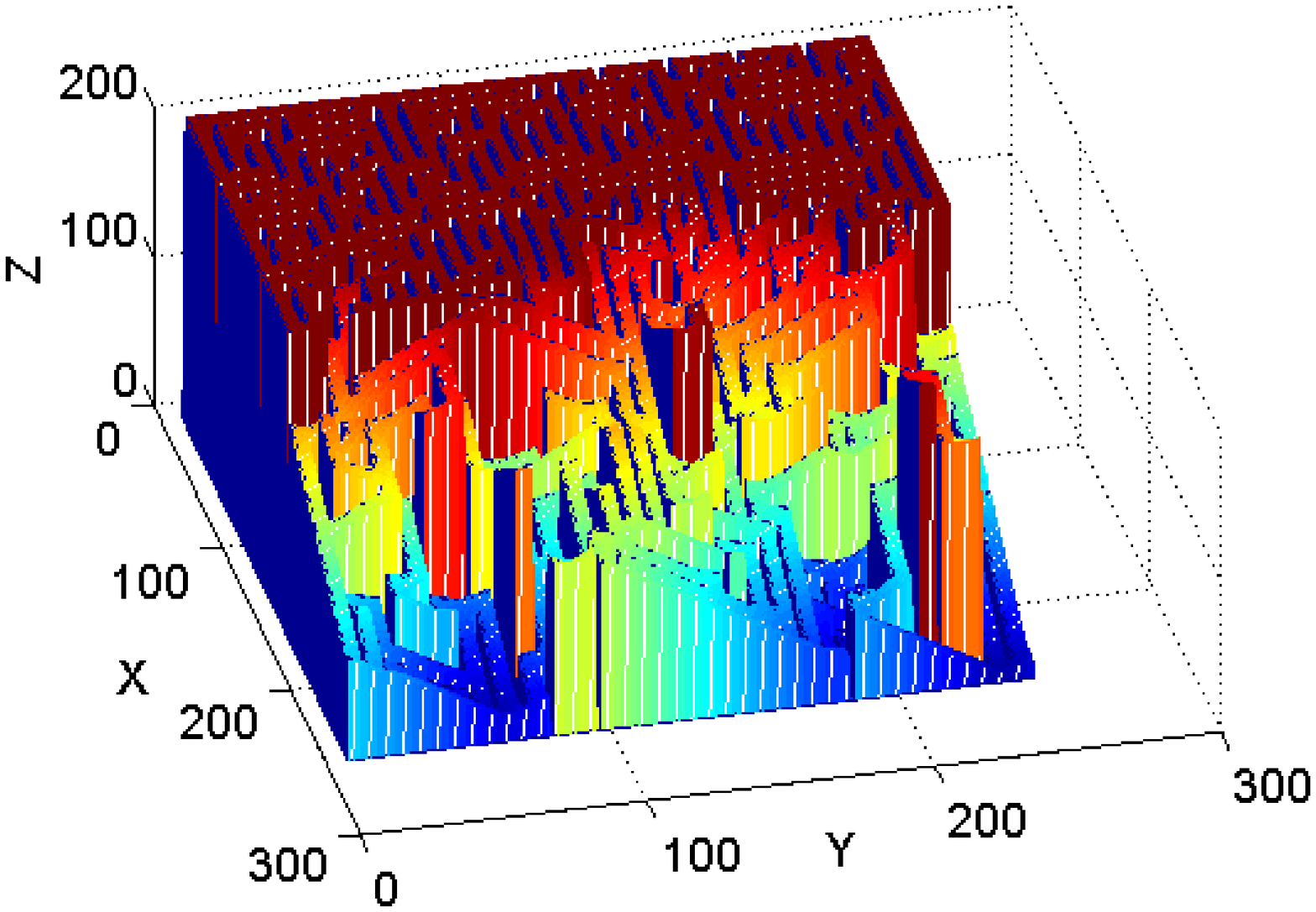}}
\subfigure[$t=0.03$ s]{\includegraphics[scale=0.3]{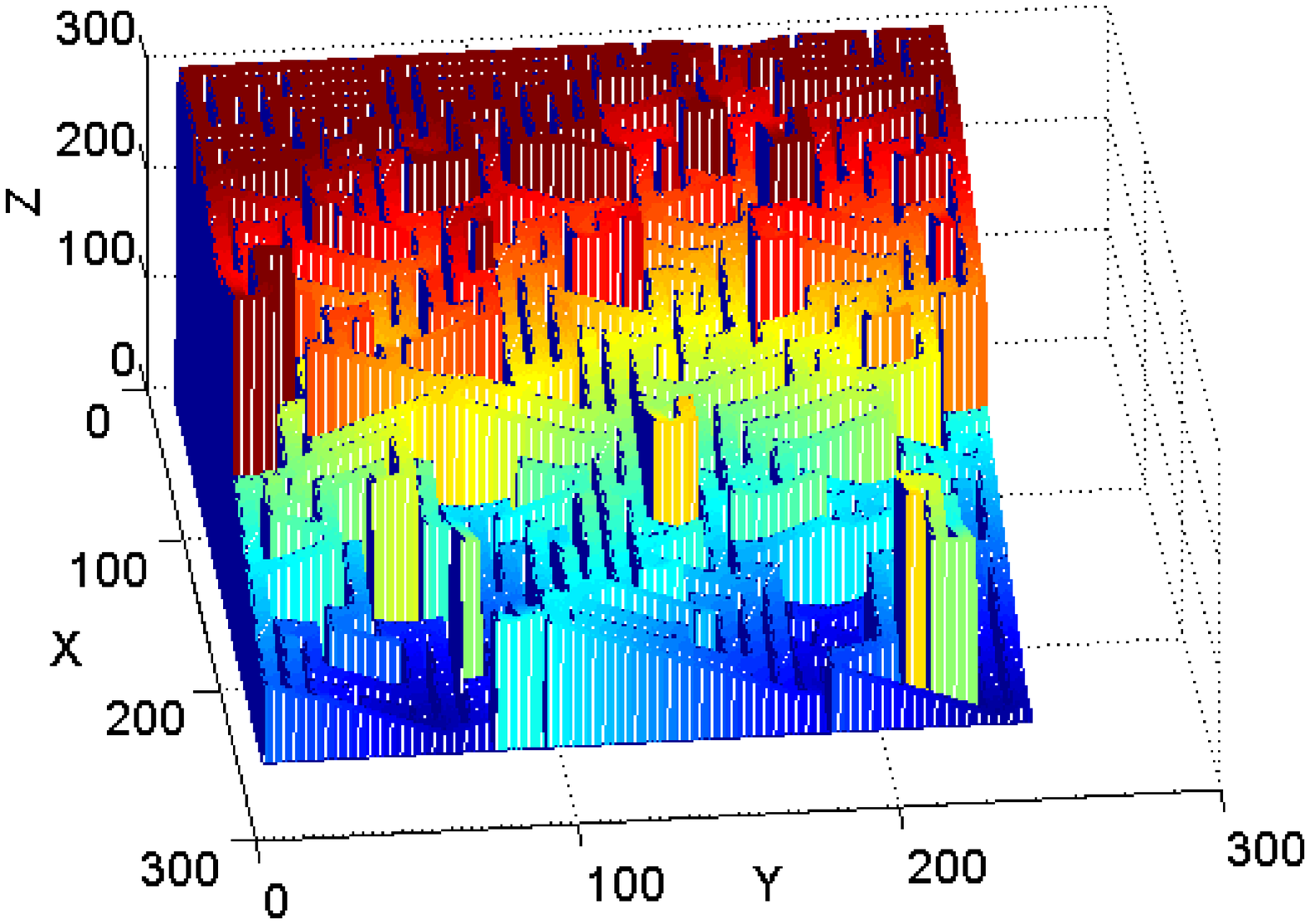}}
\subfigure[$t=0.04$ s]{\includegraphics[scale=0.3]{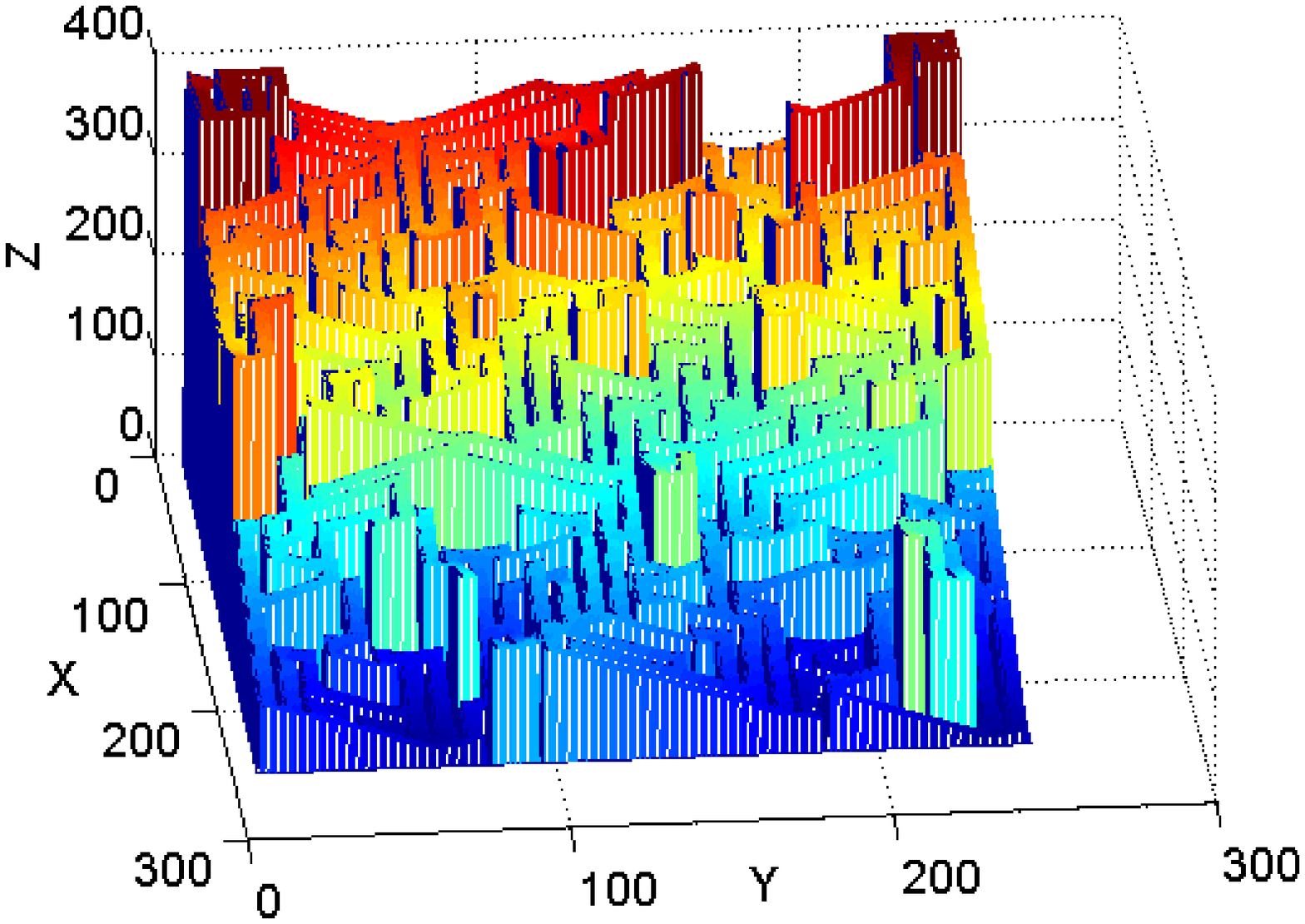}}
\subfigure[$t=0.05$ s]{\includegraphics[scale=0.3]{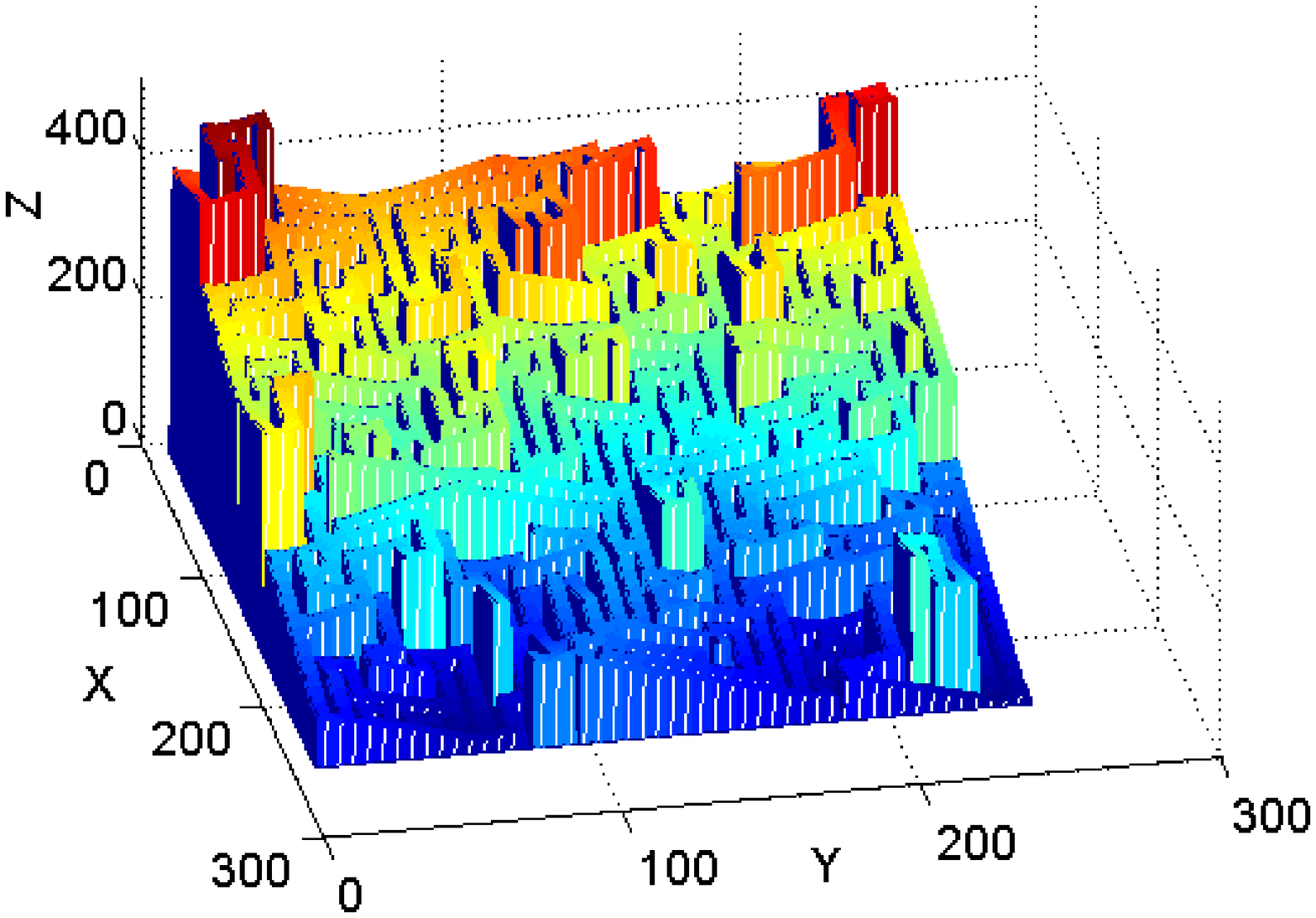}}
\caption{Transient behavior of the state values of each node on the
graph associated with the map in Fig. \ref{fig.ex21} via biased
min-consensus. \label{fig.ex22}}
\end{figure*}

\subsection{Application to Maze Solving: A Large Scale Graph with 43826 Nodes}

In this section, we further present the result about solving a maze
problem via biased min-consensus protocol (\ref{nn.biased}), which
further substantiate the efficacy and scalability of the biased
min-consensus. Consider the maze shown in Fig. \ref{fig.ex21} with
the initial position marked by a rectangle and destination positions
marked by circles. There are $254\times254$ pixels in the graph. The
pixels in the graph is divided into two classes, i.e., the obstacle
pixels with the black color and the free pixels with the white
color. We employee biased min-consensus protocol (\ref{nn.biased})
to solve the maze navigation problem. Each pixel in the free
positions of the maze graph is viewed as a node in a network. This
forms a large-scale connected graph containing 43826 nodes. The
pixels corresponding to the feasible destination positions are
viewed as static leader nodes with the state values being 0. The
pixels corresponding to other free positions are dynamic follower
nodes.

\begin{figure}[t]\centering
\includegraphics[scale=0.5]{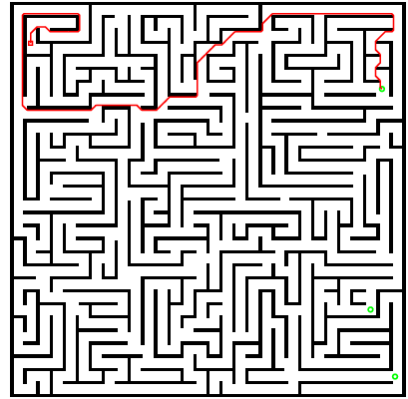}
\caption{The shortest path generated via biased min-consensus
protocol (\ref{nn.biased}) in a maze environment.\label{fig.ex23}}
\end{figure}

Using biased min-consensus protocol (\ref{nn.biased}) with parameter
$\varepsilon=1e-4$, the transient behavior of the state values of
nodes in the graph shown in Fig. \ref{fig.ex21} is shown in Fig.
\ref{fig.ex22}. The transient behavior of the state values of nodes
can be described as follows. Initially, the state values are
randomly generated and are thus unordered. Before achieving the
equilibrium, it can be seen that the state values of the nodes
corresponding to the positions far from the destination positions
evolute as traditional consensus, almost reaching the same value.
This is due to the fact that the information from the leader nodes
have not delivered to them yet. Then, with the transfer of
information, due to the effect of min-consensus protocol
(\ref{nn.biased}), the state values gradually converge to the
equilibrium. In this sense, biased min-consensus protocol
(\ref{nn.biased}) actually drives the nodes to build up gradients of
lengths of shortest paths from any position to a destination
position. Intuitively, the shortest path from any position to a
destination position is the path that follows the directions of the
fastest speed of gradient descending. The shortest path generated by
based on biased min-consensus protocol (\ref{nn.biased}) is shown in
Fig. \ref{fig.ex23}. It can be artificially checked that the
generated path is the shortest among the paths from the initial
position to all the feasible destination positions. In addition, two
videos about using biased min-consensus protocol (\ref{nn.biased})
for maze solving are available at
\url{https://www.youtube.com/watch?v=isDA0Q7LVis} and
\url{https://www.youtube.com/watch?v=fPB9-3HiSPw}. These results
further substantiae the efficacy of the biased min-consensus
protocol for solving complex shortest path problems, indicating that
we may investigate high-level intelligence from the perspective of
control.

\begin{figure}[t]\centering
\subfigure[]{\includegraphics[scale=0.16]{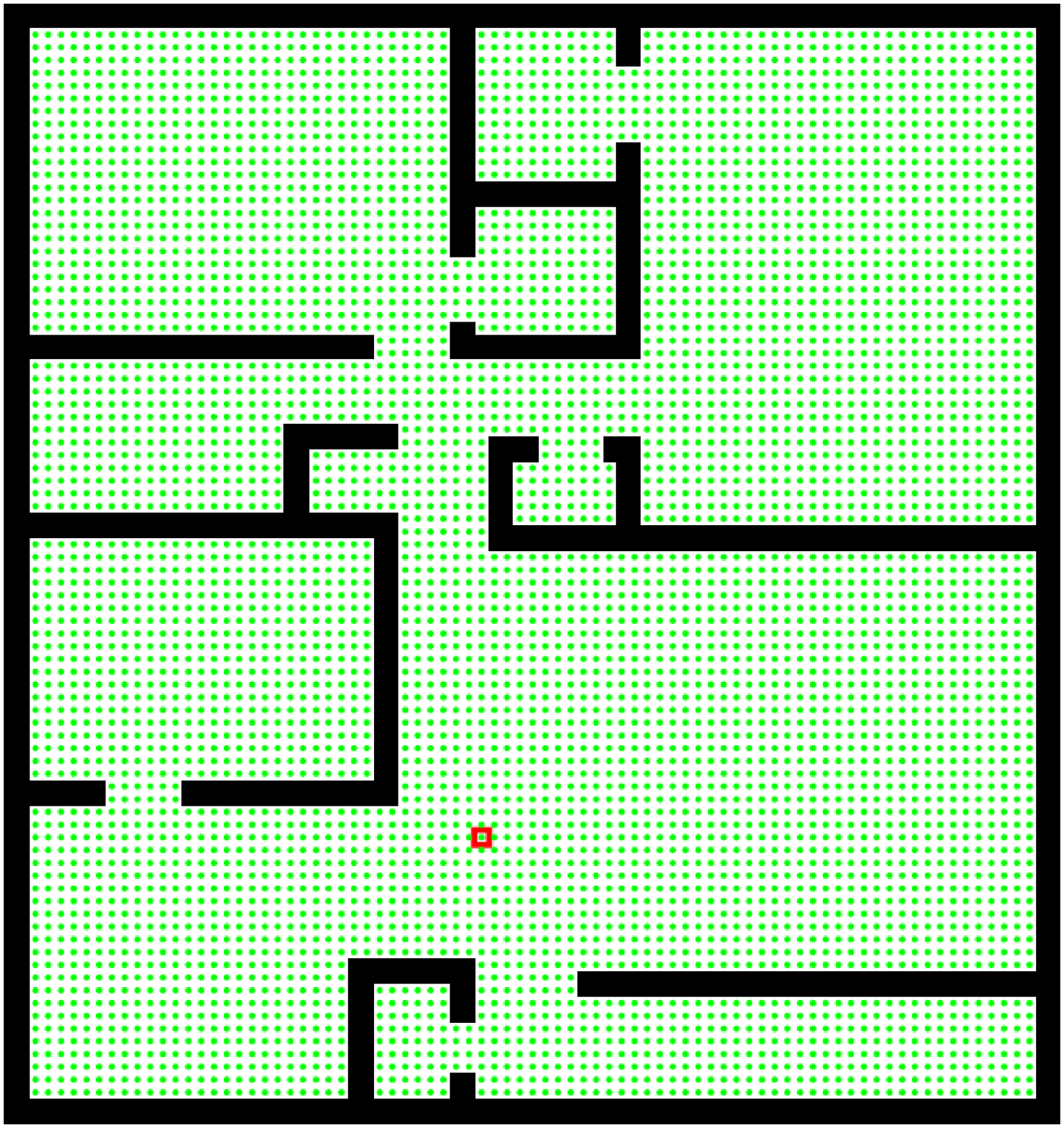}}
\subfigure[]{\includegraphics[scale=0.1627]{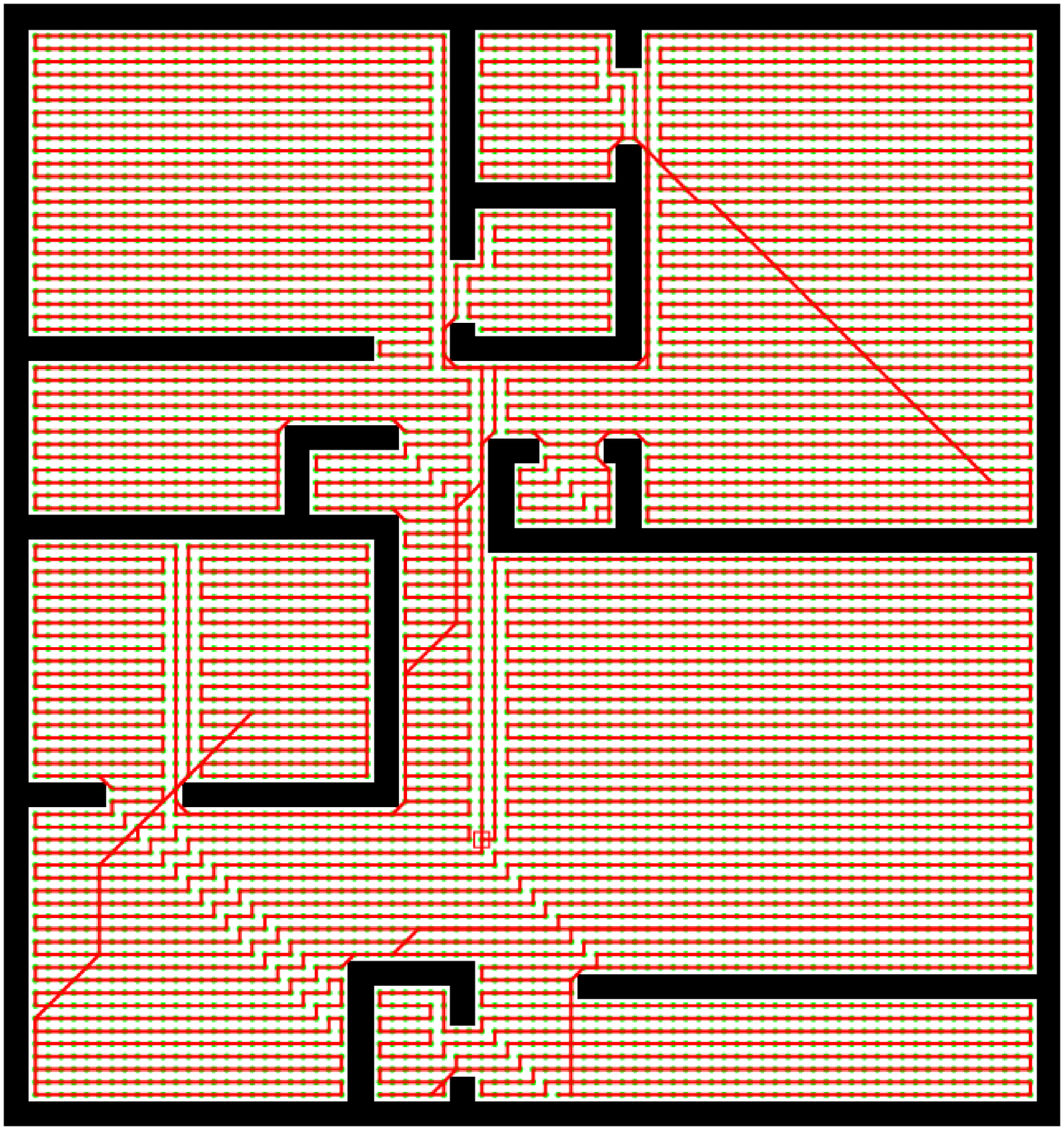}} \caption{Using
biased min-consensus protocol (\ref{nn.biased}) for complete
coverage. (a) Initial state with the initial position of a mobile
robot marked by a rectangle and areas to be covered in green. (b)
The generated moving path of the robot by using biased
min-consensus, from which we can observe that the green area has
been completely covered. \label{fig.ex31}}
\end{figure}

\subsection{Application to Complete Coverage}

In this subsection, biased min-consensus protocol (\ref{nn.biased})
is extended to solve a complete converge problem. The complete
coverage problem of mobile robots requires that a robot passes every
reachable position of the workspace \cite{psacc}. This problem is an
essential issue in cleaning robots.

According to the definition of the complete coverage problem, it can
be viewed as a extension of the shortest path problem and thus can
be solved via biased min-consensus protocol (\ref{nn.biased}). We
can treat each pixel in the free positions of the workspace as a
node of a network. In this situation, set $\mathbb{S}_1$ corresponds
to the set consisting of free (or reachable) positions that the
mobile robot has not passed. Besides, set $\mathbb{S}_2$ corresponds
to the set consisting of the positions that the mobile robot has
passed. Let $p(t)$ denotes the position of the mobile robot at time
instant $t$. The complete coverage problem can thus be solved via
the following procedure:
\begin{itemize}
\item[1)] Find the shortest path among the paths from current position $p(t)$ of the mobile
robot to all the positions in set $\mathbb{S}_1$ and drive the robot
to follow the path until it reaches the end of the path. During the
movement process of the mobile robot, remove the nodes corresponding
to the positions that the mobile robot has passed from set
$\mathbb{S}_1$;
\item[2)] If $\mathbb{S}_1$ is not empty, go to step 1); Otherwise, stop.
\end{itemize}

An example is shown in Fig. \ref{fig.ex31}. The initial state of the
complete coverage is shown in Fig. \ref{fig.ex31}(a), where the
reachable positions in the workspace is marked with small circles,
and the initial position of the mobile robot is marked with a
rectangle. By the procedures stated above, the complete coverage
result by the biased min-consensus protocol is shown in Fig.
\ref{fig.ex31}(b). As seen from this subfigure, the complete
coverage is successfully completed with lines showing the
trajectories of the mobile robot, i.e., the mobile robot has passed
each free position in the workspace. In addition, two videos about
using the biased min-consensus for complete coverage are available
at \url{https://www.youtube.com/watch?v=CCFFRfsy8CM} and
\url{https://www.youtube.com/watch?v=XkSzABQz3qw}. The results
further substantiate the efficacy of the biased min-consensus
protocol in solving the complete converge problem.

\section{Conclusions}\label{sec.6}

In this paper, we have shown that the dynamics of a biased
min-consensus protocol, as a simple evolution, can generate a
complex behavior (i.e., shortest path planning), which may trigger
our attempt to explore complex behaviors from the perspective of
consensus.  Theoretical analysis has shown that via the biased
min-consensus, the state values of the nodes on an undirected
connected graph asymptotically converge to the solution of the
shortest path problem. In addition, simulations have confirmed the
efficacy and scalability of biased min-consensus in solving shortest
path problems and revealed the potential of using biased
min-consensus for various applications, including maze solving and
complete coverage. The results obtained in this paper indicate
potential investigations on problems arising in artificial
intelligence from the perspective of consensus.

%
%
%
%
%

\end{document}